\documentstyle[12pt,preprint,prd,aps]{revtex}
\begin{document}

\def\pp{\hbox{pp}\tcar}

\draft
\begin{titlepage}
\preprint{\vbox{
\hbox{CMU-HEP-96-12}
\hbox{IASSNS-AST 96/49}
\hbox{hep-ph xxxxxxx} 
\hbox{September  1996} 
}}
\title{How Large Is the $^7{\rm Be}$ Neutrino Flux from the Sun ?}
\author{L. Wolfenstein$^a$ and P.I. Krastev$^{b}$}
\address{$^a$ Carnegie Mellon University, Pittsburgh, PA 15213}
\address{$^b$ Institute for Advanced Study\\
Princeton, NJ 08540\\}
\maketitle
\begin{abstract}
On the basis of present solar neutrino observations and relaxing the
constraints from solar models it is possible that most (or nearly all)
of the flux of electron neutrinos detected comes from electron capture
in $^7{\rm Be}$. These solutions arise from neutrino oscillations in
which $\nu_e$ $-$ $\nu_\tau$ mixing suppresses high energy $\nu_e$ and
$\nu_e$ $-$ $\nu_\mu$ mixing suppresses low energy $\nu_e$ as
qualitatively suggested from some SO(10) grand unified models. The
importance of future observations is emphasized.
\end{abstract}
\end{titlepage}

\section{Introduction}
\label{intro}

Solar neutrinos provide the only direct evidence of the nuclear
reactions that are believed to be the primary source of energy inside
the sun. Solar model calculations \cite{BP95} show that the energy is
produced mainly in the $pp$ cycle which yields three major neutrino
sources: (1) $pp$ neutrinos with energies below 420 KeV, (2) neutrinos
from electron capture in $^7{\rm Be}$ with a line spectrum and a main
energy of 862 KeV, and (3) the much rarer $^8{\rm B}$ decay neutrinos
with a spectrum up to 15 MeV. Three types of experiments have detected
neutrinos: (1) the Kamiokande experiment \cite{KAM} using
neutrino-electron scattering sensitive only to $^8{\rm B}$ neutrinos,
(2) the Davis $^{37}{\rm Cl}$ detector \cite{CHLOR} sensitive
primarily to $^8{\rm B}$ but also to $^7{\rm Be}$ neutrinos and (3)
the two gallium experiments, GALLEX \cite{GALLEX} and SAGE
\cite{SAGE}, sensitive to all three sources but, due to the rareness
of $^8{\rm B}$ neutrinos, primarily important for the $pp$ and $^7{\rm
Be}$ sources. By combining the data from the three experiments one can
try to determine the flux from each source that reaches the earth.

In the standard analysis repeated in many papers \cite{flux} one uses
the observed Kamiokande detection rate to determine the flux of
$\nu_e$ from $^8{\rm B}$ that reaches the earth. One then applies the
so obtained constraint on the boron neutrino flux to the $^{37}{\rm
Cl}$ experiment and finds that the observed rate is explained by
$^8{\rm B}$ neutrinos allowing for little or no $^7{\rm Be}$
neutrinos. If there are no $^7{\rm Be}$ neutrinos then one can
understand the gallium results as being due primarily to $pp$ neutrinos
with flux equal to that of the standard solar model (SSM). Thus one
concludes that there must exist neutrino oscillations that almost
totally convert $^7{\rm Be}$ $\nu_e$ into another type of neutrino but
have little effect on $pp$ neutrinos. This leads to solutions for
neutrino masses and mixing, in particular, the small angle MSW
solution with $\Delta m^2 \approx 10^{-5}$ ${\rm eV}^2$ and
$\sin^22\theta \approx 10^{-2}$.

It is worth pointing out \cite{john} that without oscillations one
doesn't obtain a good fit to the data, even if all the solar neutrino
fluxes are allowed to vary assuming arbitrary non-negative values,
subject only to the luminosity constraint. In this case the minimum
$\chi^2_{\rm min} = 5.9$ and it occurs for zero CNO and beryllium
neutrinos, boron neutrino flux equal to 0.35 of its standard solar
model value of $6.62\times 10^6 ~{\rm cm}^{-2}{\rm s}^{-1}$ and for a
$pp$ neutrino flux equal to that maximally allowed by the luminosity
constraint value of $6.51 \times 10^{10} ~{\rm cm}^{-2}{\rm s}^{-1}$.

Recently a number of authors\cite{{BK},{KP},{KS},{HL}} have considered
relaxing the constraints of the SSM in order to determine what is
really known about the fluxes, at the same time allowing neutrino
oscillations. One assumes only that the neutrinos arise from known
nuclear reactions and that the presently observed luminosity is the
result of these same reactions. The neutrino oscillation parameters,
$\Delta m^2$ and $\sin^22\theta$, are varied so as to fit present
experiments with non-standard solar neutrino fluxes. It turns out that
solutions can be found even for neutrino fluxes vastly different from
the ones in the SSM.  An extreme possibility \cite{BFK}, which assumes
MSW as a solution, allows that nearly all of the solar energy arises
from the CNO bicycle of nuclear reactions and nearly all of the
$\nu_e$ observed in gallium and chlorine experiments are from
$^{13}{\rm N}$ and $^{15}{\rm O}$ decays.

Here we consider the possibility that most or nearly all the $\nu_e$
observed are from $^7{\rm Be}$, the opposite of the ``conventional''
interpretation of solar neutrino experiments. It is interesting to
note that the total detected rates in chlorine and gallium are both
very close to twice that expected in the SSM for $^7{\rm Be}$
neutrinos. To obtain this result we need an oscillation solution in
which the low-energy $pp$ and high-energy $^8{\rm B}$ neutrinos are
more suppressed than those from $^7{\rm Be}$. This can occur within
the grand-unified theory (GUT) see-saw model when the large mass in
the see-saw formula is close to the GUT scale. Then high-energy
neutrinos are suppressed by $\nu_e - \nu_\tau$ oscillations but the
low energy neutrinos could be suppressed by the $\nu_e - \nu_\mu$
oscillations \cite{LW}.

It should be emphasized that we have no reason to expect the
deviations from the SSM considered here. The main purpose of this
study is to point out how little we directly know from published solar
neutrino experiments and how much can be learned from SNO, BOREXINO
and other future experiments. The spirit of this discussion is similar
to the demonstration in \cite{BFK} that the existing experiments
cannot, if neutrino oscillations occur, rule out a CNO energy
generation scenario.

This paper is organized as follows: In section \ref{Beflux} we discuss
the range of possible beryllium neutrino fluxes assuming MSW
transitions are taking place inside the Sun. The necessary conditions
are formulated that would allow for the beryllium neutrino flux to
take on its maximum allowed by the luminosity constraint value. The
realization of these conditions in a scenario where three neutrinos
take part in MSW transitions inside the Sun is described in section
\ref{three}. The allowed parameter regions are found and the
implications for neutrino masses are discussed. In section \ref{Tc} we
describe a three-neutrino MSW solution which assumes a 5 \% higher
central temperature in the sun than in the SSM. The beryllium neutrino
flux in this solution turns out to dominate the event rates in both
gallium and chlorine experiments. In section \ref{future} the expected
event rates in the future detectors SuperKamiokande, SNO, BOREXINO,
ICARUS, HELLAZ and HERON are calculated. We also discuss some of the
distinctive features of the signals in the future solar neutrino
experiments the non-observation of which will rule out the maximum
$^7{\rm Be}$ neutrino flux solution.

\section{Beryllium Neutrino Flux}
\label{Beflux}
In this section we present in a simplified form the values of the
fluxes from each of the major neutrino sources ($pp$, $^7{\rm Be}$ and
$^8{\rm B}$) allowed by the present experimental data subject to the
luminosity constraint and assuming MSW oscillations as a solution of
the solar neutrino problem,

The luminosity constraint follows from energy conservation assuming
the amount of energy emitted by the sun is matched by the amount of
energy produced in the nuclear reactions.  It can be written as a
linear relation between the neutrino fluxes from each neutrino source
($pp$, $^7{\rm Be}$, $^8{\rm B}$ etc.) \cite{BK}.  Assuming
essentially constant solar luminosity, the present value of which has
been measured with a better than a percent accuracy, any solar model
has to satisfy this constraint. As explained in \cite{BK} the
luminosity constraint imposes upper limits on all neutrino
fluxes. When the beryllium neutrino flux reaches its maximum value,
($\Phi(^7{\rm Be}) = 3.33\times 10^{10} ~{\rm cm}^{-1}{\rm s}^{-1}$),
the $pp$ neutrino flux is reduced by a factor of approximately 0.565
(assuming the same $pep/pp$ flux ratio as in the SSM) and all the
other neutrino fluxes should be zero. However, in order to explain the
Kamiokande result, one has to account for a boron neutrino flux
roughly between 0.3 and 3 times its standard solar model value
\cite{KP}. As the boron neutrino contribution to the solar luminosity
is very small this does not significantly change the upper limit on
the beryllium neutrino flux.

In a simplified version, suitable for the discussion of the general
ideas here, the luminosity constraint can be written as:
\begin{mathletters}
\label{lumina}
\begin{equation}
f_{pp} = 1.10 - 0.083 f_{Be},\label{luminaa}
\end{equation}
\begin{equation}
f_{pp} \geq 0.087 f_{Be}.\label{luminab}
\end{equation}
\end{mathletters}
The quantities $f_\alpha$, where $\alpha$ $=$ $pp$, $^7{\rm Be}$,
$pep$, $^8{\rm B}$ etc., are the ratios of the true fluxes to the ones
predicted in the SSM: $\Phi(\alpha)/\Phi(\alpha)_{\rm SSM}$.  The
numerical coefficient in Eq.\ref{luminab} is equal to the ratio
$(\Phi(^7{\rm Be})/\Phi(pp))_{\rm SSM}$ of the $^7{\rm Be}$ to $pp$
neutrino flux in the SSM and follows from the fact that one $pp$
reaction is needed to produce a $^7{\rm Be}$ nucleus. It has been
assumed in Eq.\ref{lumina} that $f_{CNO}$ = 0.0 and $f_{pep}$ =
$f_{pp}$, i.e. that the ratio between $pep$ and $pp$ neutrino fluxes
is the same as in the standard solar model.\footnote{Note that $f_{pp}
> 1$ for $f_{Be} = 1$. This reflects the fact that, since for
simplicity we have set the CNO neutrino flux has to zero, some of the
other fluxes have to be larger than in the SSM in order to compensate
for the approximately 2 \% of the luminosity that the CNO neutrinos
account for in the SSM.}  As mentioned in the previous paragraph, from
Eq.\ref{lumina} it follows that the maximum $^7{\rm Be}$ neutrino flux
is about 6.5 times the SSM $^7{\rm Be}$ flux.

In order to find the fluxes and average survival probabilities, for
which the scenario we consider here can provide a good fit to the
data, we write the event rates in the three detectors (gallium,
chlorine and neutrino-electron scattering) as:

\begin{mathletters}
\label{eq:all}
\begin{eqnarray}
Q_{Ga} & = & f_B P_B Q_{Ga}^{B} + f_{Be}P_{Be}Q_{Ga}^{Be} + 
f_{pp}P_{pp}Q_{Ga}^{pp}, \label{eq:a} \\ 
Q_{Cl} & = & f_B P_B Q_{Cl}^{B} + f_{Be}P_{Be}Q_{Cl}^{Be}, \label{eq:b} \\ 
R_{\nu e}& = & f_B(0.855 P_B + 0.145 ). \label{eq:c}
\end{eqnarray}
\end{mathletters}

The quantities in the lefthand side of the first two equations are the
event rates in the gallium and chlorine detectors, $R_{\nu e}$ is the
ratio (measured $^8{\rm B}$ flux)/(SSM ${\rm B}^8$ flux). $P_\alpha$
is the average survival probability for neutrinos from source $\alpha$
and $Q^{\alpha}_i$ is the SSM contribution of source $\alpha$ in the
i-th radiochemical detector ($i = Ga, Cl$). Here we have neglected the
``minor'' solar neutrino sources, namely the CNO and $pep$
neutrinos. We have also assumed that the suppression factors $P_B$ are
the same for all three experiments and the factors $P_{Be}$ are the
same for chlorine and gallium. In reality these factors are different
from experiment to experiment because the experiments do not cover the
same energy range and also because the detector cross-sections are
slightly different functions of energy. These simplifications are made
in this section, but not in sections \ref{three}, \ref{Tc} and
\ref{future}, in order to make more clear the general idea of the
proposed scenario.

Equation \ref{eq:c} is based on the fact that the Kamiokande detector
is sensitive to $\nu_\mu$ and $\nu_\tau$, as well as $\nu_e$, with the
cross-section for $\nu_\mu$ and $\nu_\tau$ down by a factor of about
0.15. For $R_{\nu e} = 0.44$, as indicated by the Kamiokande result,
when $f_B$ is varied from 0.3 to about 3, the required value of $P_B$
goes from one to zero. When $f_B$ is close to its maximum allowed
value nearly all the neutrinos observed by Kamiokande are $\nu_\mu$ or
$\nu_\tau$.

When we turn to the $^{37}{\rm Cl}$ detector only the $\nu_e$ are
effective. As a result, as $P_B$ approaches zero, there is no $^8{\rm
B}$ neutrino signal and, neglecting $^{13}{\rm N}$ and $^{15}{\rm O}$
neutrinos, the detected rate must be almost entirely due to $^7{\rm
Be}$ neutrinos. This requires a $^7{\rm Be}$ rate about twice the SSM
value which in turn means that nearly all the signal in the gallium
detector is due to $^7{\rm Be}$ neutrinos.

Given the experimental results:\footnote{We have combined
quadratically the statistical and systematic errors for each
experiment.}  $Q_{Ga}$ = 74 $\pm$ 8 SNU (combined GALLEX + SAGE
result), $Q_{Cl}$ = 2.55 $\pm$ 0.25 SNU, $R_{\nu e}$ = 0.44 $\pm$
0.06, and the contributions of the individual sources to each
experiment according to the SSM \cite{BP95}, which includes helium
and heavy element diffusion, we can solve Eq.\ref{eq:all} for the
products $f_\alpha P_\alpha$, assuming different boron neutrino
fluxes. The first five columns of Table \ref{tabLin} summarize the
results of this simple calculation using the central values of the
experiments.

For the last three rows of Table \ref{tabLin} the value of
$f_{Be}P_{Be}$ is greater than unity, which means that half or more of
the detected flux in the chlorine and gallium detectors is due to
$^7{\rm Be}$ neutrinos. The last row corresponds to the extreme case
that nearly all of these fluxes is due to $^7{\rm Be}$. In order to
get the large suppression needed for the $pp$ neutrinos for such a
solution we require $P_{pp} < P_{Be}$. This can be achieved by
arranging for a non-adiabatic MSW oscillation to suppress these
fluxes.

The survival probability in the non-adiabatic regime is given to a
good approximation by a simple function of energy, namely $P =
\exp(-C/E_\nu)$, where $C = \pi\Delta m^2\sin^22\theta/4\cos2\theta 
N^{'}_e$, $N^{'}_e$ is the logarithmic derivative of the electron
density at the resonance, and $E_\nu$ is the neutrino energy.  Thus it
is not possible to have very small $P_{pp}$, e.g. $P_{pp} = 0.1$, and
$P_{Be}\simeq 1$. Using the nonadiabatic survival probability for
$^7{\rm Be}$ (0.862 MeV) and $pp$ neutrinos (assuming average $pp$
neutrino energy $E_{pp} = 0.3$ MeV) to find the relation $P_{pp} =
P_{Be}^{2.87}$, one can then use Eq.\ref{luminaa} together with the
values of $f_{pp}P_{pp}$ and $f_{Be}P_{Be}$ to determine the values of
$f_{Be}$, $P_{Be}$ and $P_{pp}$ given in the last three columns of
Table \ref{tabLin}.

The last row of Table 1 illustrates the case in which the initial
beryllium neutrino flux is close to the maximum allowed by the
luminosity constraint ($f_{Be} = 6.46$). In the next section we
explore solutions of this type in the three-neutrino oscillation
scenario. A similar large beryllium neutrino flux can be fitted with
two-neutrino oscillations \cite{BK}, but in that case the $pp$ flux is
not suppressed and the $^7{\rm Be}$ neutrino contribution does not
dominate the signal in the gallium detector.

\section{Three Neutrino Oscillation Scenario}
\label{three}
Here we discuss the three-neutrino survival probabilities and the
allowed regions that we obtain for the neutrino mixing angles and
mass-squared differences for the particular example in which the
beryllium neutrino flux has its maximum allowed by the luminosity
constraint value.

In the general case of three-neutrino oscillations \cite{threenu} the
electron neutrino survival probability is a complicated function of the
mixing angles, $\theta_{12}$ and $\theta_{13}$, of the mass-squared
differences, $\Delta m^2_{21}$ and $\Delta m^2_{31}$, and of the
density distribution along the neutrino path. The existing analytic
expressions \cite{torente} for $P(\nu_e\rightarrow\nu_e)$ in this case
are far from transparent and not easy to implement in a numerical
calculation. However, as shown in \cite{smirn}, when certain
conditions on the neutrino mass and mixing parameters are satisfied,
the three-neutrino survival probability can be written down as a
product of two-neutrino survival probabilities plus a residual term
which is suppressed by a coefficient dependent on the mixing
angles. In the limit of small mixing angles and sufficiently large
separation between the positions (in space) of the resonant transition
regions the residual term is vanishingly small. In our calculations we
have neglected this residual term. In most of the interesting
situations discussed here we have verified that it is indeed small and
doesn't contribute more than a few percent to the overall
electron-neutrino survival probability, which is sufficiently accurate
for our purposes.

Using the product of the two two-neutrino survival probabilities, for
each of which we apply the analytical description from \cite{KraP}, we
have been able to study numerically a large number of interesting
cases.  Three characteristic sets of neutrino survival probability
curves are displayed in Figure 1. The survival probabilities have been
averaged over the relevant production region (as calculated in the
SSM) of each neutrino source and are indicated with different
lines. In all three figures the transitions in the energy region to
the left of the peak (lower than the energy at the maximum) are
predominantly $\nu_e\rightarrow\nu_\mu$ and those to the right of the
peak (higher than the energy at the maximum) are predominantly
$\nu_e\rightarrow\nu_{\tau}$.  The position of the peak, its height
and width are determined by all four parameters indicated in each
panel. For the maximum $^7{\rm Be}$ neutrino flux solution the width
of the peak is not very important because the CNO neutrino flux is
negligible in this case. Changes in $\theta_{13}$ do not affect
strongly the peak, but are very important for the suppression of the
high-energy boron neutrino flux, e.g. in Fig.1b, the high energy part
of the boron neutrino flux is less suppressed than in Fig.1a.  The
curves in Fig.1a and Fig.1b at energies below about 0.8 MeV coincide
except for energies $E_\nu < 0.2$ MeV, which is below the threshold
(0.233 MeV) of the gallium experiments. This is explained by the fact
that the ascending parts of the curves represent the non-adiabatic
branches of the two-neutrino survival probabilities. The latter depend
mainly on the product $\Delta m^2\sin^22\theta$ which is the same in
both panels. In Fig.1c the flux of the low energy, $pp$ and $^7{\rm
Be}$, neutrinos are more strongly suppressed than the flux of the
intermediate energy (CNO) neutrinos. Most of the $^8{\rm B}$ neutrino
flux, essentially all above 6 MeV, is strongly suppressed.

Without a solar model that could provide the neutrino fluxes envisaged
in the scenario under discussion one might think that the survival
probabilities so described are very uncertain. However, any solar
model will have to have approximately the same density distribution,
exponential to a high degree of accuracy, if it describes a star in
hydrostatic equilibrium. Note also that helioseismology provides an
independent information about the density distribution inside a large
part of the solar interior. Although the neutrino production regions
cannot be expected to be the same as in the standard solar model,
their exact shapes are not so important as it might seem. Note that in
the energy region of the $pp$, $^7{\rm Be}$ and $^8{\rm B}$ neutrinos
the survival probability curves essentially coincide (see Fig.1). The
difference is largest in the region of the intermediate energy
neutrinos which are relatively unimportant in the scenario discussed
here.

Using the so described analytical approximation of the three-neutrino
$\nu_e$ survival probability we have performed an analysis of the
solar neutrino data from the four operating experiments in a way
similar to the standard analysis for the two-neutrino MSW
solution. Our goal is to find allowed regions in parameter space in
which it is possible to fit all the experimental results with a
maximum beryllium neutrino flux. First we choose the $pp$, $pep$,
$^7{\rm Be}$ and CNO neutrino fluxes so that they correspond to the
maximum beryllium neutrino flux allowed by the luminosity constraint
assuming the SSM $pep/pp$ flux ratio. The requirement of maximal
$^7{\rm Be}$ neutrino flux automatically sets the CNO fluxes to
zero. The values of all flux ratios ($\Phi(\alpha)/\Phi(\alpha)_{\rm
SSM}$) are given in the last line of Table \ref{explim}. We then vary
the boron neutrino flux within 0.44 to about 3 times its SSM value,
i.e. within the limits set by the Kamiokande result. In addition to
the $^8{\rm B}$ flux we vary the parameters $\Delta m^2_{21}$, $\Delta
m^2_{31}$, $\sin^22\theta_{12}$ and $\sin^22\theta_{13}$ and compute
the $\chi^2$ taking into account detection cross-section uncertainties
and experimental errors. Once we have found a value of the boron
neutrino flux ($\Phi(^8{\rm B}) = 2.9)$ for which $\chi^2$ is
sufficiently small (typically $\chi^2 < 0.7$), we then keep the
$^8{\rm B}$ flux, $\Delta m^2_{21}$ and $\sin^22\theta_{12}$ fixed and
repeat the calculation varying only $\Delta m^2_{31}$ and
$\sin^22\theta_{13}$, the two parameters corresponding to the
high-energy resonance. We repeat the same procedure for several pairs
of $\Delta m_{21}^2$ and $\sin^22\theta_{12}$ close to the ones for
which a sufficiently small $\chi^2$ has been found.  The resulting 95
\% C.L.  allowed regions (corresponding to $\chi^2 = \chi^2_{\rm min}
+ 5.99$ for two d.f.) in the two parameters, $\Delta m^2_{31}$ and
$\sin^22\theta_{13}$, are shown in Fig.2. In each of the 6 panels
$\sin^22\theta_{12} = 0.1$ and the values of $\Delta m^2_{12}$ are
indicated in the upper right corner. As noted before the results
depend primarily on the product $\Delta m^2_{21}\sin^22\theta_{12}$.
Therefore quite similar plots as those shown in Fig.2 can be obtained
for different $\Delta m^2_{21}$ and $\sin^22\theta_{12}$ the product
of which is the same as in the six panels of Fig.2.  The black dots
indicate the position of the best-fit point in each case.  The
survival probability for the best fit solution in Fig.2d is the one
displayed in Fig.1a; the corresponding event rates expected in each of
the four operating experiments are given in Table II for the best fit
solution. The resulting event rates in the four experiments in this
case correspond to mainly beryllium neutrinos contributing in all
radiochemical experiments and Kamiokande detecting only $\nu_\tau$ but
virtually no $\nu_e$ from the sun.

The solutions shown correspond to the extreme possibility of a maximum
beryllium neutrino flux. Somewhat similar solutions could be found for
less extreme cases such as those illustrated by the third and fourth
rows of Table 1. As the values of $f_B$ (and $f_{Be}$) go down the
value of $\Delta m^2_{21}$ is decreased and that of $\Delta m^2_{31}$
is increased for given values of $\theta_{12}$ and $\theta_{13}$. This
yields an increase in the peak of the survival probability curve which
corresponds approximately to $P_{Be}$.

The values of $\Delta m^2_{31}$ correspond to a $\nu_\tau$ mass of the
order a few times $10^{-3} ~{\rm eV}$ in qualitative agreement with
the see-saw formula with large mass near the GUT scale. The $\nu_\mu$
mass can then be chosen a factor of 30 or so lower in general
agreement with the assumption of a mass hierarchy, although many
detailed GUT models give a significantly larger ratio.

\bigskip

\section{High $T_c$ Scenario}
\label{Tc}
The nuclear reactions in which solar neutrinos are produced take place
in a hot plasma and the neutrino fluxes are functions of the
temperature.  The effective temperature dependences of the neutrino
fluxes in the standard solar model \cite{BP95} have recently been
estimated \cite{bulmer}:

\begin{mathletters}
\begin{eqnarray}
\phi(pp) \propto  [1 - 0.08(T_c/T_{c,SSM})^{-11}], \label{fluxpp} \\
\phi(pep) \propto (T_c/T_{c,SSM})^{-2.4}, \label{fluxpep} \\
\phi(^8{\rm B}) \propto (T_c/T_{c,SSM})^{24}, \label{fluxb8} \\
\phi(^7{\rm Be}) \propto (T_c/T_{c,SSM})^{10}, \label{fluxbe7} \\
\phi(^{13}{\rm N}) \propto (T_c/T_{c,SSM})^{24.4}, \label{fluxn13} \\
\phi(^{15}{\rm O}) \propto (T_c/T_{c,SSM})^{27.1}. \label{fluxo15} 
\end{eqnarray}
\label{eqTc}
\end{mathletters}

{}From Eq.\ref{eqTc} it is clear that in solar models with higher
central temperature (and no radically new physics) the $pp$ neutrino
flux will be reduced, whereas the $^7{\rm Be}$ and $^8{\rm B}$
neutrino fluxes will be enhanced.  As discussed in the previous
section, these changes are in the direction of the ones needed in the
scenario where the $^7{\rm Be}$ neutrinos are the major component of
the signal in the radiochemical experiments. Because of the strong
temperature dependence of most of the fluxes it turns out that it is
possible to find a solution with $^7{\rm Be}$ neutrino dominance for
$(T-T_c)/T_c = 0.05$.\footnote{In order to satisfy the luminosity
constraint we have reduced the $pp$ neutrino flux by 2.5 \% from the
value prescribed by Eq.\ref{fluxpp}. For all other fluxes we have used
values given by Eq.\ref{eqTc}.} The solution that we have found is
described in Table \ref{explimTc}. The ``best-fit'' ($\chi^2_{\rm min}
= 0.7$) neutrino oscillation parameters are $\Delta m_{21}^2 = 2\times
10^{-8}~{\rm eV}^2$, $\Delta m_{31}^2 =6.3\times 10^{-5}~{\rm eV}^2$,
$\sin^22\theta_{12} = 0.1$ and $\sin^22\theta_{13} = 4.5\times
10^{-3}$.  The survival probability averaged over the different
neutrino production regions in the sun is given in Fig.1c. The major
component of the signal in the radiochemical detectors comes from
$^7{\rm Be}$ neutrinos.  The $pp$ neutrino event rate in the gallium
detectors is about 2.5 times lower than the one due to beryllium
neutrinos. The allowed region in the $\Delta m_{31}^2 -
\sin^22\theta_{13}$ plane for fixed $\Delta m_{21}^2$ and
$\sin^22\theta_{12}$ is given in Fig.3 where the black dot indicates
the ``best-fit'' solution. 

We want to emphasize that we do not introduce any particular mechanism
to change the central temperature of the sun.  It should be noted that
the temperature dependence given in Eqs.\ref{eqTc} has been derived
from a set of models in which the temperature varied by less than 2 \%
and our extension to a change of 5 \% is not necessarily
justified. Furthermore, there exists helioseismological evidence
against such a large central temperature \cite{BPhelio}; however, our
goal has been to see what can be learned from neutrino data alone.

\section{Future Detectors}
\label{future}

The scenarios we consider here have important implications for the SNO
detector. In the set of possibilities shown in Table 1, $f_B P_B$
gives the ratio of the charged current event rate (CC) to the same
rate in the SSM without oscillations. This varies from about one-third
(the value for the standard MSW solution) to close to zero in the
extreme case. SNO can distinguish the CC events on deuterium ($\nu_e +
d \rightarrow p + p + e^-$) from the neutrino electron scattering  
because the latter are strongly forward peaked, whereas the former are
weakly backward peaked. Thus for the first time the rate of $\nu_e$
arriving from $^8{\rm B}$ will be measured. The smaller the ratio of
charged current events to those from ($\nu e$) scattering, the larger
the flux of $\nu_\tau$ or $\nu_\mu$ compared to $\nu_e$ must be. In
some of the solutions the spectrum of the charged current events may
be distorted; for example, for the case shown in Fig.1b the spectrum
is suppressed at low energies and rises at high energies relative to
the normal $^8{\rm B}$ spectrum.

SNO will also measure the rate of neutral current disintegration of
deuterium (NC) ($\nu + d \rightarrow p + n + \nu$) which is directly
proportional to $f_B$ and is independent of oscillations between
active neutrinos. For the set of possibilities shown in Table 1 as the
charged current event rate decrease the neutral current one increases,
yielding the possibility of a very large neutral current to charged
current ratio.

SuperKamiokande measures only neutrino-electron scattering from
$^8{\rm B}$ neutrinos, the overall rate of which was already determined
by the earlier Kamiokande experiment. This experiment is sensitive to
significant distortions in the spectrum of detected neutrinos. In the
extreme case we consider in which nearly all the neutrinos detected
are $\nu_\mu$ and $\nu_\tau$ there is very little or no distortion at
all. The day-night effect which could occur because of regeneration of
$\nu_e$ in the earth is not expected for the values of $\Delta m^2$
in our solutions.

Future detectors like BOREXINO \cite{BOREXINO}, ICARUS \cite{ICARUS},
HERON \cite{HERON} and HELLAZ \cite{HELLAZ} might become operational
in about 5-10 years. BOREXINO will be the first detector to directly
measure the beryllium neutrino flux. According to the standard
two-neutrino MSW solution based on the SSM, this detector is unlikely
to measure any $^7{\rm Be}$ flux above background whereas in the
solution discussed in Sec.3 the detected flux could be more than a
factor of two greater than in the SSM. HERON and HELLAZ could be the
first detectors to directly measure the $pp$ neutrino flux. It is
generally believed that the $pp$ neutrinos are little affected by MSW
oscillations but in the solutions considered here the arriving flux of
$pp$ neutrinos may be severely suppressed and the spectrum distorted.

Table \ref{futureexp} summarizes our predictions for the event rates
in future detectors in two scenarios with beryllium neutrino flux
higher than in the SSM; these are compared to the predictions of the
standard two-neutrino MSW solutions. The first line corresponds to the
solution where the $^7{\rm Be}$ neutrino flux is 6.46 times higher
than in the SSM and almost equal to the maximum allowed by the
luminosity constraint. The second line corresponds to the solution for
neutrino fluxes from a hypothetical solar model with a central
temperature higher by 5 \% than the one in the SSM. The normalized
CC/NC double ratio (${\rm (CC/CC_{SSM})/(NC/NC_{SSM})}$) in SNO can be
obtained from the column corresponding to the SNO(CC) event rate by
dividing by $f_B$ (2.9 for the first line and 3.22 for the second
line). The results for BOREXINO include only the neutrino-electron
scattering events originating from $^7{\rm Be}$ neutrinos and include
the contribution from ($\nu_\mu e$) and/or ($\nu_\tau e$) scattering.
The high event rate in BOREXINO in the maximum beryllium neutrino
scenario could be a clear signal of oscillations because it requires a
contribution from $\nu_\mu$ or $\nu_\tau$ neutrinos in order to be
consistent with the measured $\nu_e$ rates in gallium and chlorine.

\section{Conclusions}

The ultimate goal of solar neutrino experiments is to learn about the
nuclear reactions in the sun and about possible neutrino
oscillations. Given the limited present data the standard approach has
been to use a standard model for the sun and deduce neutrino
oscillation parameters. These yield the result that practically no
$\nu_e$ from $^7{\rm Be}$ have been detected.

Here we relax the constraints on the fluxes from the SSM, keeping
intact only the luminosity constraint, and demonstrate that the
present data are consistent with the possibility of nearly all the
$\nu_e$ detected in radiochemical detectors being from $^7{\rm Be}$
and that nearly all the signal in the Kamiokande detector in this case
could arise from $\nu_\mu$ or $\nu_\tau$ scattering from
electrons. The neutrino oscillation parameters required are
qualitatively consistent with neutrino masses and mixing suggested by
some grand unified theories.

The major goal of this exercise is to emphasize how little is directly
known about the solar neutrino flux arriving on earth and the
importance of future experiments, particularly SNO, BOREXINO, and
HELLAZ and HERON, which will be the first experiments to directly
measure the $\nu_e$ flux from $^8{\rm B}$ decay, and the $^7{\rm Be} +
p$ and $pp$ reactions.

\section*{Acknowledgments}
The work of L.W. was partially supported by the U.S. Department of
Energy Contract No. DE-FG 02-91 ER40682. The work of P.K. was
partially supported by NSF grant \#PHY92-45317. We are thankful to
J. Bahcall and K.S. Babu for usefull discussions.  P.K. acknowledges
support by the Theory Group at Fermilab for their summer visitor
program during which part of this work was completed.


\newpage
\begin{table}
\caption{
Fluxes and survival probabilities depending on the assumed boron
neutrino flux (first column) which satisfy the constraints from the
existing data (columns two to five) which are then combined with the
luminosity constraint and the specific energy dependence of the
survival probability for non-adiabatic MSW transitions to get the last
three columns.}

\begin{tabular}{l c c c c c c c}
$f_B$ & $P_B$ & $f_B P_B$ & $f_{Be}P_{Be}$ & $f_{pp}P_{pp}$ & $f_{Be}$ 
& $P_{Be}$ & $P_{pp}$ \\ \hline 
1.0 & 0.34~ & 0.34~ & 0.0085 & 0.98~ & 0.0089 & 0.96 & 0.89~ \\ 
1.5 & 0.17~ & 0.26~ & 0.51~~ & 0.72~ & 0.58~~ & 0.88 & 0.69~ \\ 
2.0 & 0.088 & 0.17~ & 1.02~~ & 0.47~ & 1.31~~ & 0.77 & 0.48~ \\ 
2.5 & 0.036 & 0.091 & 1.52~~ & 0.22~ & 2.48~~ & 0.61 & 0.24~ \\ 
2.9 & 0.008 & 0.023 & 1.92~~ & 0.017 & 6.46~~ & 0.30 & 0.031 \\ \hline
\end{tabular}
\label{tabLin}
\end{table}

\begin{table} 

\caption{Event rates in the operating experiments 
(three neutrino oscillations, maximum $^7{\rm Be}$ neutrino flux).
The values in the table are in SNU for chlorine and gallium and the
ratio (observed $^8{\rm B}$ flux)/(SSM $^{\rm B}$ flux) is given for
Kamiokande. The neutrino mass and mixing angles are given in the text
and correspond to the black dot in Fig.2d.  The last line in the table
gives the ratios of the assumed neutrino fluxes to the ones in the SSM
\protect\cite{BP95}.}

\begin{tabular}{l c c c c c c c c}
Experiment & $pp$ & $^8{\rm B}$ & $^7{\rm Be(0.862)}$ & 
$^7{\rm Be(0.384)}$ & $^{13}{\rm N}$ & $^{15}{\rm O}$ & $pep$ & total \\ 
\hline
Gallium  & 2.6  & 0.30 & 70.2 & 0.70 & 0.0 & 0.0  & 0.65  & 74.5 \\
Chlorine &  -   & 0.14 & 2.38 &  -   & 0.0 & 0.0  & 0.05  & 2.57 \\
Kamioka  &  -   & 0.44 &   -  &  -   &  -  &   -  &   -   & 0.44 \\ 
$f_\alpha$& 0.56 & 2.9  & 6.46 & 6.46 & 0.0 & 0.0  & 0.56  & \\ 
\end{tabular}
\label{explim}
\end{table}

\begin{table} 

\caption{Event rates in the operating experiments 
(three neutrino oscillations, $T_c = 1.05 T_c^{SSM}$).  The values in
the table are in SNU for chlorine and gallium and the ratio (observed
$^8{\rm B}$ flux)/(SSM $^{\rm B}$ flux) is given for Kamiokande. The
neutrino mass and mixing angles are given in the text. The last line
in the table gives the ratios of the assumed neutrino fluxes to the
ones in the SSM \protect\cite{BP95}.}

\begin{tabular}{l c c c c c c c c}
Experiment & $pp$ & $^8{\rm B}$ & $^7{\rm Be(0.862)}$ & $^7{\rm Be(0.384)}$ 
& $^{13}{\rm N}$ & $^{15}{\rm O}$ & $pep$ & total \\ \hline
Gallium  & 13.7  & 1.00 & 32.4 & 0.55 & 5.82 & 15.0  & 1.83  & 70.3 \\
Chlorine &  -   & 0.18 & 1.10 &  -   & 0.19 & 1.10  & 0.13  & 2.70 \\
Kamioka  &  -   & 0.48 &   -  &  -   &  -  &   -  &   -   & 0.48 \\ 
$f_\alpha$& 0.90 & 3.22  & 1.63 & 1.63 & 3.29 & 3.75  & 0.90  & \\ 
\end{tabular}
\label{explimTc}
\end{table}

\begin{table} 
\caption{Ranges of event rates in the future experiments for: a) three
neutrino oscillations, maximum $^7{\rm Be}$ neutrino flux; b) $T_c =
1.05T_{c,SSM}$, c) small mixing-angle two-neutrino MSW solution (SMA)
and d) large mixing-angle two-neutrino MSW solution (LMA).  Each entry
in the table represents the range of the ratio of the predicted event
rate in the corresponding scenario to the event rate assuming SSM
fluxes and no oscillations.  The central value of the ratio
corresponds to ``best-fit'' neutrino oscillation parameters in the
relevant case. The neutrino mass and mixing angles are given in the
text for the first two solutions and in \protect\cite{BK} for the SMA
and LMA solutions.}

\begin{tabular}{l c c c c c }
scenario   & SuperK & SNO(CC)  & BOREXINO  & ICARUS & HELLAZ/HERON \\ \hline
maximum $^7{\rm Be}$ & $0.45^{+0.12}_{-0.01}$ & $0.019^{+0.10}_{-0.012}$ & $2.88^{+0.06}_{-0.39}$ & $0.019^{+0.10}_{-0.012}$ & $0.20^{+0.0}_{-0.0}$ \\ 
high $T_c$  & $0.51^{+0.24}_{-0.02}$ & $0.012^{+0.15}_{-0.005}$ & $1.04^{+0.0}_{-0.20}$ & $0.010^{+0.14}_{-0.002}$ & $0.41^{+0.0}_{-0.01}$ \\ 
SMA ($2\nu$) & $0.41^{+0.19}_{-0.13}$ & $0.32^{+0.23}_{-0.16}$ & $0.22^{+0.50}_{-0.01}$ & $0.34^{+0.23}_{-0.18}$ & $0.96^{+0.04}_{-0.31}$ \\ 
LMA ($2\nu$) & $0.34^{+0.09}_{-0.06}$ & $0.22^{+0.23}_{-0.06}$ & $0.59^{+0.13}_{-0.14}$ & $0.22^{+0.11}_{-0.06}$ & $0.73^{+0.06}_{-0.08}$ \\ 
\end{tabular}
\label{futureexp}
\end{table}

\begin{figure}
\bigskip\bigskip

\caption[]{
Neutrino survival probabilities averaged over the relevant neutrino
production regions (dotted line $pp$, short-dash-dotted line $^7{\rm
Be}$, full line $^8{\rm B}$, short-dashed line $^{13}{\rm N}$,
long-dashed line $^{15}{\rm O}$, long-dash-dotted line $pep$), are
shown as a function of energy. The neutrino mass-squared differences
and mixing angles are indicated in each panel. The type of the
dominant transition ($\nu_e \rightarrow \nu_\mu$ or $\nu_e \rightarrow
\nu_\tau$) is indicated on both sides of the peak in the first
panel.\label{survP}}

\bigskip\medskip

\caption[]{ Allowed regions at 95 \% C.L. in the mass-mixing plane
($\Delta m_{31}^2 - \sin^22\theta_{13}$). The other two neutrino
oscillation parameters are fixed: for each of the six panels
$\sin^22\theta_{12} = 0.1$ and the values of $\Delta m^2_{21}$ are
given in the upper right corner of each panel. The point in the
mass-mixing plane where the $\chi^2$ is minimum is indicated with a
black dot. The boron neutrino flux is equal to $2.9 \Phi(^8{\rm
B})_{\rm SSM}$, the $pp$ neutrino flux is 56 \% of the SSM value, and
the beryllium neutrino flux is equal to 99.9 \% of the maximum allowed
by the luminosity constraint value ($3.33 \times 10^{10} {\rm
cm}^{-2}{\rm s}^{-1}$).  }

\bigskip\medskip

\caption[]{
Allowed region at 95 \% C.L. in the mass-mixing plane ($\Delta
m_{31}^2 - \sin^22\theta_{13}$) corresponding to neutrino fluxes in a
solar model with $T_c = 1.05 T_{c,SSM}$. The neutrino oscillation
parameters $\Delta m^2_{21}$ and $\sin^22\theta_{12}$ are fixed and
their values are indicated in the panel. The black dot corresponds to
the point in the mass-mixing plane where the $\chi^2$ is minimum.  The
ratio of the neutrino fluxes to those in the SSM are given in the last
line of Table \ref{explimTc}. }

\end{figure}


\newpage
\begin{thebibliography}{20}
\bibitem{BP95} J. Bahcall and M. Pinsonneault, 
Rev. Mod. Phys. {\bf 67}\, 1 (1995); for a detailed description of the
physics involved in solar modeling see J.N. Bahcall, {\it Neutrino
Astrophysics (Cambridge University Press, Cambridge, England, 1989)}.

\bibitem{KAM}KAMIOKANDE Collaboration, Y. Suzuki,
Nucl. Phys. B (Proc. Suppl.) {\bf 38}, 54 (1995); K. S. Hirata {\it et
\ al.}, Phys.  Rev. D {\bf 44} 2241 (1991).


\bibitem{CHLOR} B. T. Cleveland {\it et al.},
Nucl. Phys. B (Proc. Suppl.) {\bf 38}, 47 (1995); R. Davis,
Prog. Part. Nucl. Phys. {\bf 32}, 13 (1994).



\bibitem{GALLEX} GALLEX Collaboration, P. Anselmann {\it et al.},
Phys. Lett. B {\bf 327}, 377 (1994); {\bf 342}, 440 (1995); {\bf
 357}, 237 (1995).

\bibitem{SAGE} SAGE Collaboration, G. Nico {\it et al.}, in
 {\it Proceedings of the XXVII International Conference on High Energy
Physics}, Glasgow, Scotland, 1994, edited by P. J. Busse\ y and I.
G. Knowles (Institute of Physics, Bristol, 1995), p. 965;
J. N. Abdurashitov {\it et al.}, Phys. Lett. B {\bf 328}, 234 (1994).

\bibitem{flux} J.N. Bahcall and H. Bethe, Phys. Rev. {\bf 47}, 
1298 (1993); P. Rosen and W. Kwong, Phys. Rev. Lett. {\bf 73}, 369
(1994); J.N. Bahcall, Phys. Lett. B {\bf 338}, 276 (1994); S. Parke,
Phys. Rev. Lett. {\bf 74}, 839 (1995); N. Hata and P. Langacker,
Phys. Rev. D {\bf 52}, 420 (1995). 


\bibitem{john} J.N. Bahcall, Invited talk at the symposium on The
Inconstant Sun, Naples, Italy, March 18, 1996; to be published in
Memorie della Societa, eds. G. Cauzzi and C. Marmolino.


\bibitem{BK} J.N. Bahcall and P.I. Krastev, Phys. Rev. D {\bf 53}, 
4211 (1996).

\bibitem{KP} P.I. Krastev and S.T. Petcov, Phys. Rev. D {\bf 53}, 
1665 (1995).

\bibitem{KS} P.I. Krastev and A.Yu. Smirnov, Phys. Lett. B {\bf 338}, 
282 (1994).

\bibitem{HL} N. Hata and P. Langacker, Phys. Rev D {\bf 50}, 632 (1994).

\bibitem{BFK} J.N. Bahcall, M. Fukugita and P.I. Krastev, 
Phys. Lett. B {\bf 374}, 1 (1996).

\bibitem{LW} L. Wolfenstein, Phys. Rev. D {\bf 45}, R4365, (1992).

\bibitem{threenu} T.K. Kuo, J. Pantaleone, Phys.Rev.D35:3432,1987,
S. Toshev, Phys.Lett. B {\bf 185}, 177, (1987); Erratum: Phys. Lett. B
{\bf 192}, 478,(1987); S.T. Petcov and S.Toshev, Phys.Lett. B, {\bf
187}, 120, (1987); S.T. Petcov, Phys. Lett. B, {\bf 214}, 259 (1988).

\bibitem{torente} E. Torrente Lujan, Phys. Rev. D {\bf 53}, 4030,
(1996).

\bibitem{smirn} A.Yu. Smirnov, Sov.J.Nucl. Phys. {\bf 46}, 672
(1987) (Yad. Fiz. {\bf 46}, 1152 (1987)).

\bibitem{KraP} P.I. Krastev and S.T. Petcov, 
Phys. Lett. B {\bf 207}, 64 (1988).

\bibitem{bulmer} J.N. Bahcall and A. Ulmer, Phys. Rev. D {\bf 53},
4202, (1996).

\bibitem{BPhelio} J.N. Bahcall and M. Pinsonneault, ``Are Standard
Solar Models Reliable?'', preprint.

\bibitem{SuperK}  M. Takita, in {\it Frontiers of Neutrino Astrophysics},
edited by Y. Suzuki and K. Nakamura (Universal Academy Press, Tokyo,
1993), p. 147; T. Kajita, {\it Physics with the SuperKamiokande
Detector}, ICRR Report 185-89-2 (1989).

\bibitem{SNO} H. H. Chen, Phys. Rev. Lett. {\bf 55}, 1534 (1985); G.
Ewan {\it et al.}, Sudbury Neutrino Observatory Proposal, SNO-87-12
(1987);A. B. McDonald, in Proceedings of the Ninth Lake Louise Winter
Institute, edited by A. Astbury {\it et al.} (World Scientific,
Singapore, 1994), p. 1.

\bibitem{BOREXINO}  C. Arpesella {\it et al.}, BOREXINO proposal, Vols. 1
and 2, edited by G. Bellini, {\it et al.} (Univ. of Milano, Milano,
1992); R. S. Raghavan, Science {\bf 267}, 45 (1995).  

\bibitem{ICARUS} A First 600 Ton ICARUS Detector Installed at the 
Gran Sasso Laboratory, addendum to proposal LNGS-94/99 I\&II, preprint
LNGS-95/10 (1995); J. N. Bahcall, M. Baldo-Ceolin, D.  Cline and
C. Rubbia, Phys. Lett. B. 178 (1986) 324.

\bibitem{HERON}  S. R. Bandler {\it et al.},
Journal of Low Temp. Phys. {\bf 93}, 785 (1993);
R. E. Lanou, H. J. Maris, and G. M. Seidel, Phys. Rev. Lett. {\bf 58},
2498 (1987).

\bibitem{HELLAZ} G. Laurenti {\it et al.}, in {\it Proceedings
of the Fifth International Workshop on Neutrino Telescopes, Venice,
Italy, 1993}, edited by M. Baldo Ceolin (Padua University, Padua,
Italy.


\end{thebibliography}
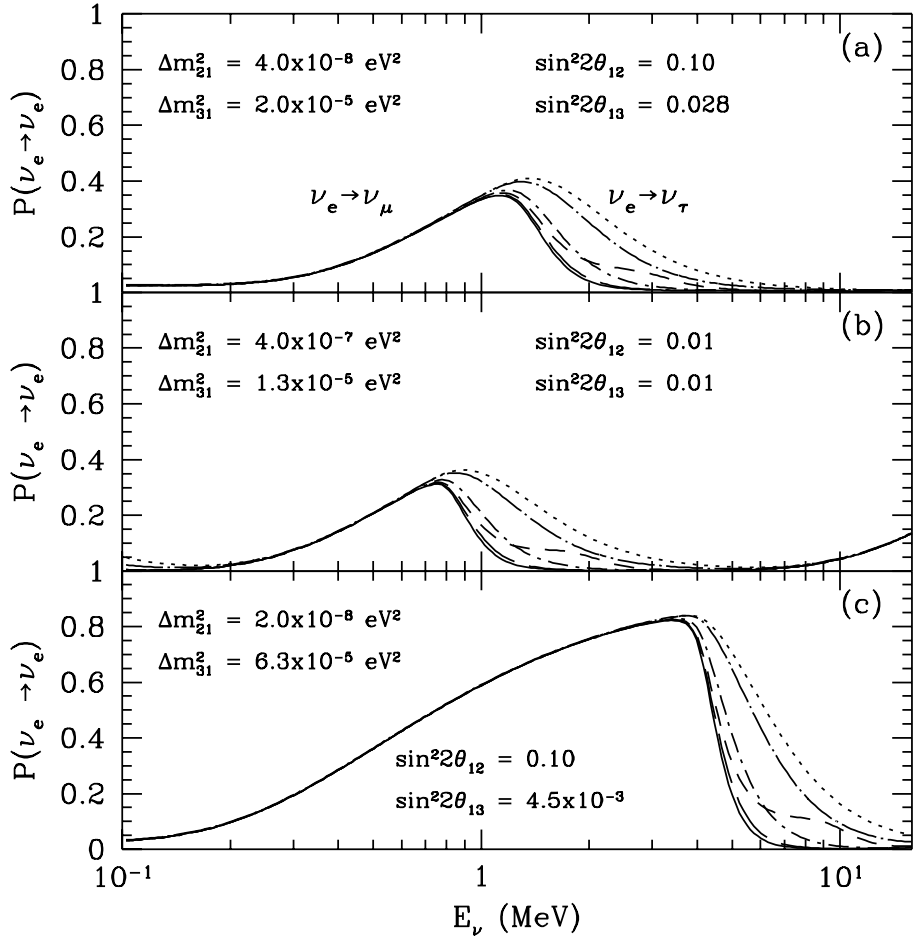
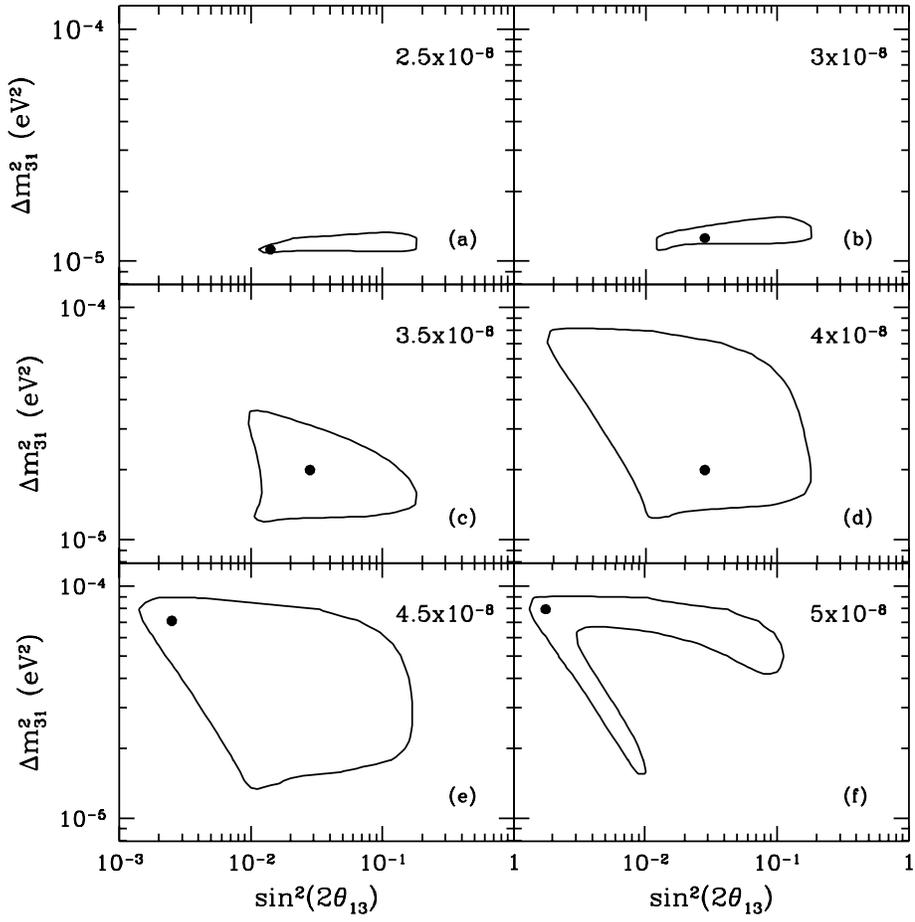
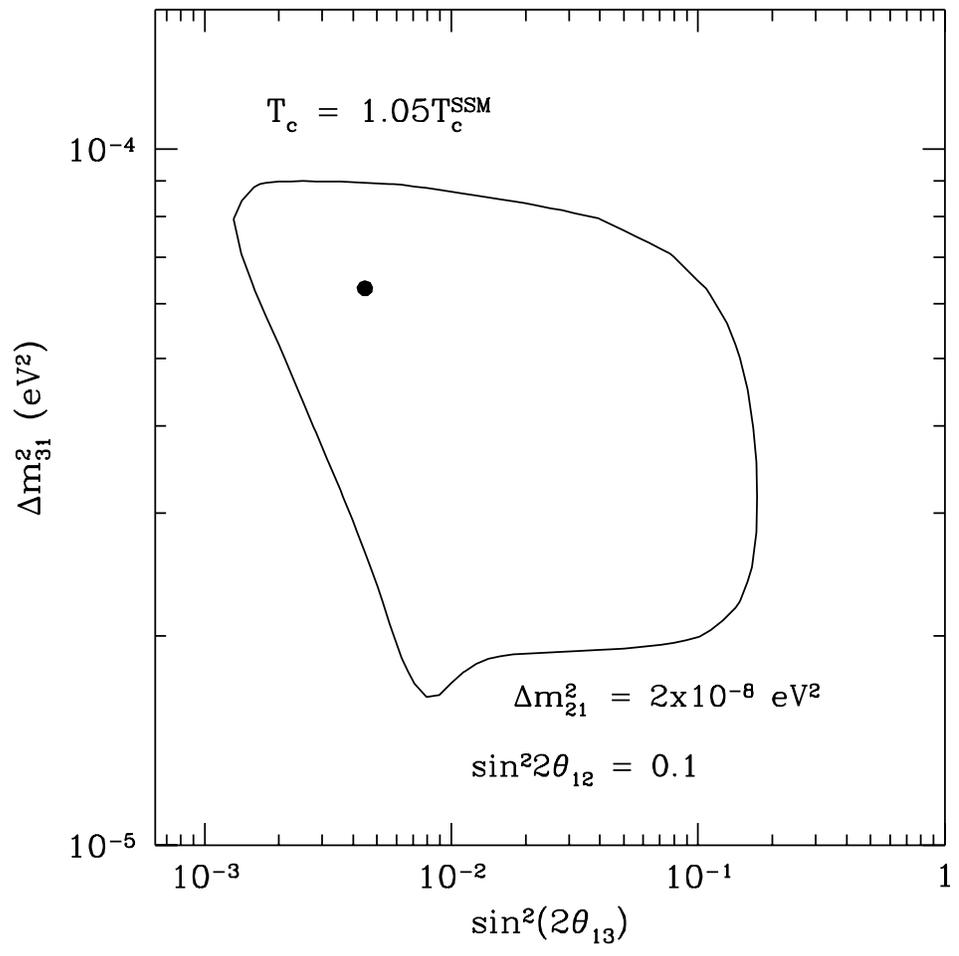
\end{document}